\documentstyle[11pt,newpasp,,twoside,epsf]{article}
\begin{document}
	\title{
Infrared Surveys for AGN}

	\author{
Harding E. Smith\altaffilmark{1}}

	\affil{
Center for Astrophysics \& Space Sciences 
and
Department of Physics,
University of California, San Diego,
La Jolla, CA 92093-0424,
USA}

\altaffiltext{1} {
also, Infrared Processing and Analysis Center,Caltech/JPL, 
                   Pasadena, CA 91125}

	\begin{abstract} 
From the earliest extragalactic infrared studies AGN have shown themselves
to be strong infrared sources and IR surveys have revealed new populations
of AGN.  I briefly review current motivations for AGN surveys in the
infrared and results from previous IR surveys.  The Luminous Infrared
Galaxies, which in some cases house dust-enshrouded AGN, submillimeter
surveys, and recent studies of the cosmic x-ray and infrared backgrounds
suggest that there is a population of highly-obscured AGN at high redshift.
ISO Surveys have begun to resolve the infrared background and may have
detected this obscured AGN population.  New infrared surveys, particularly
the SIRTF Wide-area Infrared Extragalactic Legacy Survey ({\it SWIRE}),
will detect this population and provide a platform for understanding the
evolution of AGN, Starbursts and passively evolving galaxies in the context
of large-scale structure and environment.
\end{abstract}

	\section{
Motivation: Active Galaxies in the Infrared} 

There are a number of motivations for infrared studies of AGN and for
carrying out surveys for AGN at infrared wavelengths:

\begin{enumerate}
\item AGN unification models require a dusty, molecular obscuring screen
      	(torus?) which is expected to emit in the infrared.

\item Cosmic X-Ray Background models require a population of highly
	obscured AGN which may be detectable in the infrared.

\item Infrared/Starburst emission is a frequent companion to AGN activity;
	IR emission originates from dust which is either associated with 
	the active  nucleus ({\it e.g.} the PG QSO sample; Sanders 
	{\it et al.} 1989) or with circumnuclear 
	starburst emission as in the classical Seyfert galaxies NGC 1068
	\& NGC 7469.

\item Luminous Infrared Galaxies frequently harbor AGN cores and may be 
	a stage in the development of AGN from the merger of gas-rich
	galaxies.
\end{enumerate}
	\noindent

	\subsection{
AGN Unification Models}

The AGN Unification models (\cite{A93}) which have been popular for some
time, and which have garnered considerable observational support, require an
optically-thick, dusty obscuring screen, frequently assumed to be toroidal
(\cite{PK}; \cite{GD}), which obscures the central engine and broad
emission-line region from equatorial lines of sight.  In this scenario
broad-line objects (Type 1 AGN -- Sy1 galaxies and classical QSOs) are
objects viewed near the pole of the torus, while narrow-line objects (Type
2 AGN -- Sy 2 galaxies) are viewed edge-on.  Direct evidence for the
existence of this obscuring screen is largely from studies of scattered
broad-line emission in Sy 2 systems, but claims have been made that the
mid-infrared spectra of some nearby galaxies match various torus models
(\cite{PK}; \cite{GD}).  A critical point regarding this torus is that 
{\it it will be optically thick at all wavelengths from soft x-rays through
the mid-infrared.}  Even mid-infrared diagnostics may not reveal a highly
obscured AGN.  Hard x-rays and VLBI radio imaging are currently
required to peer through the veil of obscuration.

	\subsection{
The Cosmic X-Ray and Infrared Backgrounds}

Recent models for the Cosmic X-Ray Background (CXB) require a population of
highly obscured, perhaps even compton-thick, AGN which emit primarily in
the hard x-ray region in order to fit the x-ray background spectrum
(\cite{Co95}; \cite{Gi01}).  This population increases in density with
redshift. These AGN are expected to re-emit the absorbed radiation at
infrared wavelengths providing a link between the CXB and the Cosmic
InfraRed Background (CIRB).  Recent measures of the CIRB with COBE
(\cite{Ha98}; \cite{Pu96}) show that more than half of the cosmic energy
density (excluding the CMB) comes out in the infrared.  An important
question is then the relative importance of accretion energy due to AGN
compared with stellar nucleosynthesis.  Deep surveys with Chandra and
XMM-Newton are resolving the CXB (\cite{Ho01}; \cite{Ha01}; \cite{Ro02})
and ISO (\cite{E02}) has begun to resolve CIRB.  New Surveys with SIRTF
will convincingly determine relationship between the faint x-ray and 
infrared populations revealing a great deal about the history of our Universe.

	\subsection{
Luminous Infrared Galaxies}

The most luminous galaxies in the Local Universe are Luminous Infrared
Galaxies (LIGs) which emit the vast majority of their radiant power in
the far-infrared between about 40--120$\mu$m.  These are gas-rich
systems which are in the late stages of collisions or mergers. 
Extrapolation from the properties of lower luminosity Starburst
galaxies suggests that the LIGs should be active star-forming systems
(see Sanders \& Mirabel 1996 for a review).  The LIGs also show many
characterisatics of AGN and their luminosities reach values comparable
to those of luminous QSOs.  Much effort has been focused on whether
LIGs are powered principally by Starburst or AGN activity, although
both types of activity are almost certainly present.  The discussion
has been framed around a scenario proposed by Sanders {\it et al.}
(1988) in which a merger of gas-rich disk galaxies stimulates a
massive nuclear Starburst which in turn feeds a coalescing AGN core in
the galaxy nucleus.  As the AGN turns on, radiation pressure drives
out the shroud of dust, revealing a nascent quasar.  The goal must be
not only to understand the dominant source of energy in LIGs, but to
understand the relationship between Starburst and AGN activity and
other galaxy characteristics, and to place them into an evolutionary
context.  

We have for some time been using VLBI techniques to attempt to understand 
the power sources of LIGs and to place them in an evolutionary context.
Analysis of a complete LIG sample (\cite{SLL98}) suggests that many LIGs
may be interpreted as intense Starbursts like Arp 220 (Figure 1), but a
nearly equal number must house AGN cores, as in the case of Mrk 231,
whose nuclear structure implies a recent ignition ($t << 10^6 yr$) for
AGN activity.  Evidence is strong that a significant number of LIGs
house obscured AGN activity and schematics are consistent with this
Starburst-to-AGN scenario.

	\begin{figure}
	\plotfiddle{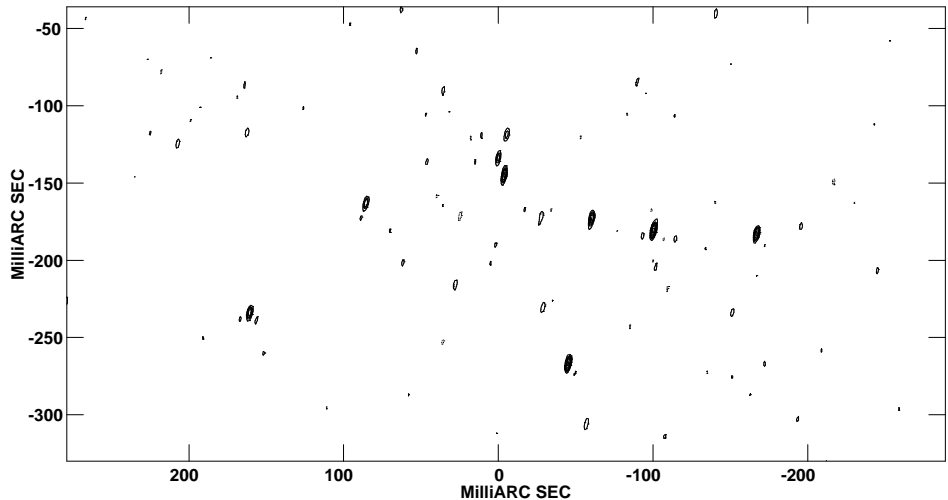}{5.8cm}{270}{50}{50}{-200}{225}
	\caption{
18 cm VLBI image of the W Nucleus of Arp 220 from Smith {\it etal.}
(1998).  Over a dozen unresolved sources are interpreted as luminous 
radio supernovae in an intense starburst.
}
	\end{figure}
\begin{figure}
\plotfiddle{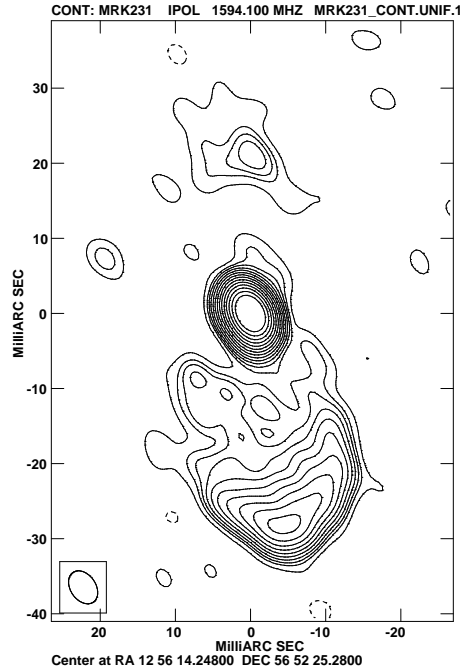}{7.9cm}{0}{40}{40}{-150}{-55}
\caption{18 cm VLBI image of the central 100pc of Mrk 231, interpreted as a
nascent QSO; age $<< 10^6$yr.}
\end{figure}

Of particular interest are the extremely luminous high-z galaxies detected
in recent SCUBA and other submillimeter surveys (\cite{I00}).  The SEDs of
LIGs in the submillimeter have the unique characteristic that the positive
K-correction offsets cosmological dimming for redshifts from $z \sim 1$ --
10 such that it is equally easy (or difficult) to detect infrared galaxies
at 850$\mu$m over a range of high redshifts.  Although only a handful of
redshifts are available for submm sources owing to the large error circles
for submm sources and the optical faintness of the small number of
identified galaxies, existing redshifts confirm that the submm population
lies at redshifts $1 < z < 3$.  The existence of these sources implies a
population of very luminous galaxies at very early epochs, with
concommittant rapid evolution.  The star-formation rates inferred from the
luminosities of these systems are not easily produced in CDM galaxy
formation models. In analogy with local LIGs, the submillimeter sources may
be candidates for nascent QSOs at high-redshift.

	\section{
Active Galaxies with IRAS}

One of the most striking results from IRAS is that the infrared galaxy
population evolves more rapidly than the optical galaxy population
(\cite{Lons90}) with a form similar to that for QSOs.
Selection of Luminous IR Galaxies associated with FIRST radio sources
(\cite{St01}) or by FIR/Optical ratio (\cite{Sm02}; \cite{Lons02}) have led
to the detection of IRAS Faint Sources up to $z \sim 1$ with a small number
of ``hyperluminous'' or lensed sources with redshifts in excess of 1
({\it e.g.} \cite{RR}).

IRAS showed that the infrared spectra of QSOs are essentially similar, with
a broad 'infrared bump' from about 2$\mu$m to 1 mm, with a significant
fraction (10--50\%) of the bolometric luminosity emitted at infrared
wavelengths.  This infrared emission is almost certainly thermal dust
emission, but the location of this warm dust remains uncertain
(\cite{San89}) possibly associated with the outer edges of the accretion
disk, the torus, or circumnuclear star formation.

Most of the known Sy 2 galaxies were discovered by IRAS, but the
sensitivity of IRAS has limited limits our census of ``Type 2 AGN'' to $z <
0.2$.  As Padovani (1998) has previously stressed, at $z < 0.2$ AGN Type 2
outnumber Type 1, suggesting that there are a large number of Sy 2 galaxies
and Type 2 QSOs yet to be discovered at higher redshift.  The expectation
from our local census and from the CXB models suggests that an important
task of future IR AGN Surveys will be detecting the high-redshift AGN 2
population, or explaining its absence in terms of AGN evolution.

	\section{
2MASS Quasars}

The 2MASS AGN surveys have been reviewed by R. Cutri (2002; this volume)
and the characteristics of that population will be discussed only briefly.
The surface density of 2MASS AGN is of the order of 0.5 deg$^{-2}$ with a
space density comparable to that found in low-redshift optical-to-x-ray
selected samples.  The colors and polarization of 2MASS QSOs suggests that 
these are {\it reddened} rather than intrinsically red QSOs and the ratio of
Type 1 AGN to Type 2 is about 2:1 --- the obscured QSO 2's are not being 
found in 2MASS.

The color selection applied to detect 2MASS QSOs is $(J-K) > 2$ and Cutri
{\it et al.} argue that these QSOs are sufficiently red to have been missed
in previous optical surveys.  Analysis by the SDSS and 2MASS teams (Ivezic
{\it et al.} 2002; this volume)
suggests, however, that the incompleteness of the SDSS to such red QSOs is 
less than 10\%.  The 2MASS AGN are faint in x-rays with a wide range in
hardness-ratio (\cite{W02}) suggesting that the faintest and hardest
sources may be highly obscured and could provide the missing x-ray
population required to explain the CXB.

	\section{
ISO AGN studies and Surveys}

	\subsection{
Mid-Infrared Diagnostics for AGN}

ISO has shown that mid-infrared spectroscopy is a powerful tool for source
classification.  In a series of papers (\cite{Lu98},\cite{Tr01}) the
mid-infrared molecular band strengths haver been used to discriminate
between AGN-dominated (7.7$\mu$m line to continuum $< 1$) and
Starburst-dominated (7.7$\mu$m to continuum $> 1$) LIGs.  These results
suggest that the majority of LIGs with $log\,L_{fir} < 12 (L_\odot)$, the
``ULIGs'', are Starburst-dominated, with an increasing AGN-fraction with
increasing $L_{fir}$.  Comparison with optical spectroscopy (\cite{LVG})
shows that the infrared classifications are consistent with classical
optical excitation methods and suggests that infrared galaxies with {\it
LINER} spectra are Starbursts (Veilleux 2002; this volume).

A somewhat different direction has been taken by Clavel {\it et al.}
(2000) who compared the strengths of the mid-infrared bands among AGN
types.  In their sample there is a clear distinction between Sy 1 galaxies,
showing low $W_\lambda(7.7\mu m)$, and Sy 1.5-2 galaxies with higher values
of $W_\lambda(7.7\mu m)$.  The 7.7$\mu$m {\it luminosities} of the Sy
galaxy types are comparable, however.  They interpret this result in terms
of the classical AGN-torus model, suggesting that the warm AGN continuum
from the inner torus is extinguished in the edge-on Sy 2 systems but
visible in the Sy 1 systems, viewed face-on, whereas the extended,
Starburst-related mid-infrared features are visible in both types of
galaxies.

Taken together these results underscore the caveat discussed above, that
AGN activity in compact, highly-obscured infrared galaxies may remain
hidden even at mid-infrared wavelengths.

	\subsection{
ISO Deep Surveys}

A number of deep surveys were undertaken with the ISO satellite at
wavelengths of 7, 15, 90, and 170$\mu$m.  These are reviewed in detail by
Taniguchi (2002; this volume) and the results will only be summarized here.
The principal result from these surveys is the continuing high surface
densities of infrared galaxies as ISO pushed to lower flux-densities --- as
low as $\sim 10\mu$Jy at the lower wavelengths --- requiring continued
steep evolution with redshift.  There are a number of models with varying
prescriptions for the evolving population (\cite{AF01}; \cite{RR01}; 
\cite{X01}).  The models of Xu {\it et al.}, based upon a local complete 
24$\mu$m sample, require luminosity evolution rates as high as $L \propto
(1+z)^{4.2}$ combined with density evolution, $\rho \propto (1+z)^2$, up to
$z \sim 1$ for the Starburst population to match the $log\,N-log\,S$
relation; the number of LIGs at $z \sim 1$ is thus estimated to be
approximately 40 times higher that in the local Universe (\cite{AF01}).
The Xu {\it et al.} model has pure luminosity evolution for the AGN
population, $L \propto (1+z)^{3.5}$, whereas the ``normal'' galaxy
population evolves as $L
\propto (1+z)^{1.5}$.

With about 2000+ galaxies resolved by ISO and over 400 spectroscopic
observations, AGN account for about 10\% of the identified sources in the
ISO Surveys.  The ratio of AGN type 1 to type 2 in these surveys is about
unity, at variance with the IRAS results at higher flux-density.

Elbaz {\it et al.} (2002) have analyzed the deep ISO 15$\mu$m counts and 
estimated that the IR galaxies detected by ISO to a 15$\mu$m flux-density,
$S_{15\mu m} > 50\mu$Jy, contribute over half of the CIRB and that AGN may
contribute, at most about 20\% of the IR background.  Fadda {\it et al.} (2002)
have combined the deep ISOCAM/XMM-Newton data from the HDFN and Lockman
Hole with the brighter ELAIS S1/BeppoSAX data to estimate the AGN
contribution th mid-IR surveys and the CIRB.  Using x-ray emission as an
indicator of AGN activity, Fadda {\it et al.} estimate that 15-20\% of the
mid-infrared emission in Lockman and HDFN originates from AGN.  The 
detection (AGN) fraction of 15$\mu$m sources increases with x-ray energy
from 30\% below 2keV to over 60\% above 5keV, as might be expected if
the x-ray background is produced by obscured, high-column sources.  Again,
the fraction of the CIRB attributed to AGN-accretion energy is estimated to
be less than 20\%.

Further analysis of red x-ray luminous galaxies 
($L_x \sim 10^{43}$--$10^{45}erg/s$)
in the same sample (\cite{AF02}) suggests that these may be highly-obscured
AGN as predicted by CXB background models.  The mid-infrared SEDs of these
sources are well reproduced by model spectra of obscured QSOs with 
$\tau_{0.3\mu m} \sim 30$--40 and the ratio of Type 1-to-Type 2 AGN is
1:3 in agreement with predictions.

The areas surveyed remain small, less than a few hundred square arcminutes
to $S_{15\mu m} < 100\mu$Jy, and statistics are restricted to small numbers
with a few tens of confirmed AGN.  The lower sensitivities at the longer
wavelengths require the above analyses to employ template SEDs for
estimating the contributions of ISO sources at the peak of CIRB near
140$\mu$m.  With the wide range of mid-to-far-IR SEDS observed in the local
Universe ({\it e.g.} the Far-IR/Mid-IR ratio may differ by an order of
magnitude between a ``typical Starburst'' like M82 and a LIG such as Arp
220) these intriguing results remain tentative.

	\section{
SWIRE: The SIRTF Wide-area InfraRed Extragalactic Survey}

For {\it SIRTF}, the last of its {\it Great Observatories}, NASA has
selected a set of Legacy Programs, designed to be major surveys of general
interest to the Astronomical community and to be carried out in the first
year of the SIRTF mission.  The Survey data will be distributed to the
community in time for use in preparation of {\it General Observer
Proposals} with no proprietary period for Legacy data.

The SIRTF Wide-area InfraRed Extragalactic Survey ({\it SWIRE}, Dr. Carol
Lonsdale, P.I.) is the largest of the six SIRTF Legacy Surveys 
(851 hours), surveying approximately 67 square degrees in all 7 SIRTF
imaging bands.  A current description of the SWIRE Survey is given on 
the SWIRE WebPages: 
{\it http://www.ipac.caltech.edu/SWIRE}. Table 1 lists the Survey 
sensitivities.

	\begin{table}
	\caption{
SWIRE Sensitivity Limits (est. 5$\sigma$)}
	\begin{tabular}{rrccrrc}
	\tableline
\multicolumn{3}{c}{IRAC} & & \multicolumn{3}{c}{MIPS} \\
$\lambda$ & Sensitivity & Resolution &&$\lambda$ & Sensitivity & Resolution\\
\tableline
   3.6$\mu$m &  7.3$\mu$Jy & 0.9$''$ & &  24$\mu$m &  0.45mJy & 5.5$''$\\
   4.5$\mu$m &  9.7$\mu$Jy & 1.2$''$ & &  70$\mu$m &  2.75mJy & 16$''$\\
   5.8$\mu$m & 27.5$\mu$Jy & 1.5$''$ & & 160$\mu$m &  17.5mJy & 36$''$\\
   8.0$\mu$m & 32.5$\mu$Jy & 1.8$''$ & &           &          &\\
\tableline
\tableline

\end{tabular}\end{table}

The Survey will cover seven high-latitude fields, selected to be the
most transparent, lowest background fields in the sky.  The fields,
covering between 5 and 15 sq. deg. include previously well-known IR
extragalactic survey fields (e.g. Lockman and the ELAIS ISO Survey 
Fields) and x-ray fields (Chandra Deep South and XMM Large Scale 
Survey) are shown in Table 2.
	\begin{table}
	\caption{
SWIRE Survey Fields}
	\begin{tabular}{lllrl}
	\tableline
Field & \multicolumn{2}{c}{Center (J2000)} & \ Area     & Background \\
      & \ RA        & \ Dec                & (sq deg) & (MJy/Sr) \\
\tableline
ELAIS S1   &  00$^h $38$^m$ 30$^s$ & $-$44$^\circ$ 00$^\prime$ & 14.8 & 0.42 \\
XMM-LSS    &  02$^h $21$^m$ 00$^s$ & $-$05$^\circ$ 00$^\prime$ &  9.3 & 1.3\\
Chandra-S  &  03$^h $32$^m$ 00$^s$ & $-$28$^\circ$ 16$^\prime$ &  7.2 & 0.46 \\
Lockman    &  10$^h $45$^m$ 00$^s$ &  +58$^\circ$ 00$^\prime$ & 14.8 & 0.38 \\
Lonsdale   &  14$^h $41$^m$ 00$^s$ &  +59$^\circ$ 25$^\prime$ &  6.9 & 0.47 \\
ELAIS N1   &  16$^h $11$^m$ 00$^s$ &  +55$^\circ$ 00$^\prime$ &  9.3 & 0.44 \\
ELAIS N2   &  16$^h $36$^m$ 48$^s$ &  +41$^\circ$ 02$^\prime$ &  4.5 & 0.42 \\
\tableline
\tableline
\end{tabular}\end{table}

The SWIRE science goal is to enable fundamental studies of galaxy
evolution in the infrared for 0.5 $ <\, z\, <$3:

	\begin{itemize}
\item evolution of star-forming and passively evolving galaxies in the
context of structure formation and environment.

\item spatial distribution and clustering of evolved galaxies, Starbursts,
\& AGN.

\item the evolutionary relationship between galaxies and AGN and the
contribution of AGN accretion energy to the cosmic backgrounds.
	\end{itemize}

Galaxy evolution models which match the IRAS/ISO galaxy counts at all
wavelengths from 7--100$\mu$m as well as the CIRB (\cite{X01}) predict that
SWIRE will detect of the order of 2 million galaxies --- spheroids and
evolved stellar systems with IRAC, and active star-forming systems with
MIPS.  SWIRE will also detect about 25,000 classical AGN, and an unknown
number, perhaps several times as many, dust-enshrouded AGN.

Recent estimates of the ``Universal Star-formation History ({\it SFH})''
(\cite{CS99}) suggest that the bulk of cosmic evolution occurs between
redshifts, $0.5 < z < 3$, the redshift interval for which SWIRE is
optimized.  The median redshift is predicted to be, $\left< z\right> \sim
1$, where many estimates find a peak in the {\it SFH}; luminous infrared
galaxies will be detected by {\it SWIRE} out to $z \sim 3$.  Previous
estimates of the {\it SFH} have varying, frequently large and uncertain
corrections for extinction.  {\it SWIRE} will directly measure the total
star-formation rates as a function of redshift and environment over this
critical range of time and redshift.

A key element in the {\it SWIRE} Survey design is to enable 
galaxy evolution studies in the context of large-scale structure/environment.
One of the {\it SWIRE} Survey fields covers the deep survey areas of the
XMM-LSS Survey (\cite{MP01}) so that the infrared galaxy census may be
directly tied to the presence of rich x-ray clusters to $z > 1$.  {\it
SWIRE} will sample several hundred, 100 Mpc scale co-moving volume cells
enabling a variety of large-scale structure measures from
correlation functions, power spectra, and counts-in-cells to direct
comparison with model calculations.  SWIRE's measures of the star-formation
as a function of environment will be important input for CDM simulations
which have been exceedingly
successful in simulating the development of structure in the early
Universe, but perhaps less so in simulating galaxy evolution within 
that structure owing to the complexity of the physics of star formation
({\it e.g.} Kay {\it et al.} 2002).

Of more direct importance to this Conference, the similarities of AGN SEDs
in the mid-far Infrared suggests that {\it SWIRE} will be unbiased with
respect to AGN types and ages, enabling a complete census of AGN out to
redshifts greater than 1.  Although the detection rates should be unbiased,
the similarity between the SEDs of obscured AGN to those of 
Starbursts, and the extreme optical depths will make {\it identifying}
the obscured AGN population very challenging.  Low-frequency radio surveys
will, of course, identify radio-loud AGN, but these make up only 10--15\%
of the AGN population.  For this reason the XMM-LSS
Survey, along with current and planned deeper surveys in hard x-rays will
be vital to identifying {\it SWIRE} AGN.

	\subsection{
Supporting Observations}

An aggressive program of ground-based optical, near-infrared and radio 
observations is planned in support of the SWIRE Survey and we are actively
pursuing other programs with HST, Chandra, XMM and Galex.  As already
described Chandra and XMM Surveys will be important for discovering the
obscured AGN population, if it exists.  SWIRE has entered into cooperation
with the Galex team  so that the SWIRE fields will be included in
the Galex Deep Survey.

The SWIRE Optical-Near Infrared goal is to obtain moderate-depth optical
multi-band ($g' \sim 25.7$, $r' \sim 25$, $i' \sim 24$; Vega magnitudes,
5$\sigma$ detection for a $2''$ galaxy) data for the entire Survey area.
At these limits we expect to detect approximately $2/3$ of SWIRE sources
detected by both MIPS and IRAC.  The ELAIS N1, N2 fields have already been
imaged to somewhat shallower limits ($r' \sim 24$) as part of the INT Wide
Field Survey and efforts continue to push deeper in the optical and into
the near infrared as part of the UK SWIRE Program (ISLES and UKIDSS
projects respectively).  An extensive program for observations of ELAIS S1
is being undertaken at ESO.  Optical and near-infrared imaging of the
Lockman, Lonsdale and CDFS fields are being undertaken at KPNO and CTIO
with the Mosaic cameras and FLAMINGOS infrared imager.

Two major SWIRE radio surveys are planned.  The median
20cm flux density predicted for SWIRE Starburst galaxies is 
$\sim 43\mu$Jy --- to faint to survey the entire area to this depth.
We have therefore planned a deep pencil-beam VLA Survey (F. Owen, PI)
and an extended shallow VLA survey (J. Condon, PI):
	\begin{itemize}
\item SWIRE Lockman Deep Survey --- 3$\mu$Jy rms @ 20cm; $\alpha = 10^h\,46^m$
	$\delta = +59^\circ\,01^\prime$; 30$^\prime$ VLA primary beam.  The
	Deep VLA Survey is nearly completed and data analysis is just 
	beginning.

\item Cosmic Windows VLA Survey --- 50$\mu$Jy rms @ 20cm in the combined
	fields of SWIRE, Galex and XMM-LSS which are accessible to the
	VLA.  This Survey is being
	proposed for the next VLA large survey program.  
	\end{itemize}

	\section{Benediction
}
The {\it SWIRE} Legacy Survey is a community Survey; the large dataset
which is being accumulated reflects the synergies which 
between the Legacy program and other community surveys.  With a couple of
million galaxies and several tens of thousands of AGN, many with redshift
estimates and SEDs from x-ray to radio, the SWIRE database will be released
to the community through IPAC's InfraRed Science Archive.  We hope that
{\it SWIRE} will provide a rich datamine for the entire community and will
provide answers to many of the questions posed here.  If
you have projects that you would like to do with {\it SWIRE} data, pleas
visit the {\it SWIRE} WebPages and/or contact
one of the team members.

It is a pleasure to thank the local organizers, especially Areg Mickaelian
and Ed Khachikian, and the scientific organizing committee for my second
stimulating visit to beautiful Armenia.  This research was
supported by the US NASA.


\begin{thebibliography}{}
	\bibitem[Antonucci 1993]{A93}
Antonucci, R. 1993, \araa, 31, 473.
	\bibitem[Clavel {\it et al.} 2000]{Cl00}
Clavel {\it et al.} 2000, \aap, {\bf 357}, 839.
	\bibitem[Comastri {\it et al.} 1995]{Co95}
Comastri, A., Setti, G., Zamorani, G., \& Hasinger, G. 1995, 
	\aap, {\bf 296}, 1.
	\bibitem[Elbaz {\it et al.} 2002]{E02}
Elbaz, D., {\it et al.} 2002, \aap, {\it in press}. (astro-ph/0201328)
	\bibitem[Fadda {\it et al.} 2002]{F02}
Fadda, D., {\it et al.} 2002, \aap, {\it in press}. (astro-ph/0111412)
	\bibitem[Franceschini {\it et al.} 2002]{AF02}
Franceschini, A., {\it et al.} 2002, \aap, {\it in press}. (astro-ph/0111413). 
	\bibitem[Franceschini {\it et al.} 2001]{AF01}
Franceschini, A., {\it et al.} 2001, \aap, {\bf 378}, 1. 
	\bibitem[Gilli {\it et al.} 2001]{Gi01}
Gilli, R., Salvati, M. \& Hasinger, G. 2001, \aap, {\bf 366}, 407.
	\bibitem[Granato \& Danese 1994]{GD}
Granato, G. \& Danese L., 1994, \mnras, {\bf 268}, 235.
	\bibitem[Hasinger {\it et al.} 2001]{Ha01}
Hasinger, G., {\it et al.} 2001 \aap, {\bf 365}, 45.
	\bibitem[Hauser {\it et al.} 1998]{Ha98} 
Hauser, M., {\it et al.} 1998, \apj, {\bf 508}, 25.
	\bibitem[Hornschemeier {\it et al.} 2001]{Ho01}
Hornschemeier, A., {\it et al.} 2001, \apj, {\bf 554}, 742.
	\bibitem[Ivison {\it et al.} 2000]{I00}
Ivison, R., {\it et al.}, 2000, \mnras, {\bf 315}, 209.
	\bibitem[Kay {\it et al.} 2002]{K02}
Kay, S., Pearce, F., Frenk, C., \& Jenkins, A. 2002, \mnras, {\bf 330}, 113.
	\bibitem[Lonsdale {\it et al.} 1990]{Lons90}
Lonsdale, C. J, {\it et al.} 1990, \apj, {\bf 358}, 60.
	\bibitem[Lonsdale {\it et al.} 2002]{Lons02} 
Lonsdale, C., Hurt, R., \& Smith, H. E.,  \& Xu, C.
		2002, \apj, {\it in preparation}.
	\bibitem[Lutz {\it et al.} 1998]{Lu98}
Lutz, D., {\it et al.} 1998, \apjl, {\bf 505}, L103.
	\bibitem[Lutz, Veilleux \& Genzel 1999]{LVG}
Lutz, D., Veilleux, S. \& Genzel, R. 1999, \apjl, {\bf 517}, L13.
	\bibitem[Padovani 1998]{Pa98}
Padovani, P. 1998 in {\it New Horizons from Multi-Wavelength Sky Surveys},
	ed. B. McLean, D. Golombek, J. Hayes, \& H. Payne, (Kluwer), p. 257.
	\bibitem[Pier \& Krolik 1993]{PK}
Pier, E. \& Krolik, J. 1993, \apj, {\bf 418}, 673.
	\bibitem[Pierre 2001]{MP01}
Pierre, M. 2001 in {\it Where's the Matter?}, eds. L. Tresse \& M. Treyer,
	{\it in press}. (astro-ph/0111242) 
	\bibitem[Puget {\it et al.} 1996]{Pu96}
Puget, J-L., {\it et al.} 1996, \aap, {\bf 308}, 5.
	\bibitem[Rosati {\it et al.} 2001]{Ro02}
Rosati, P. {\it et al.} 2002, \apj, {\bf 566}, 667.
	\bibitem[Rowan-Robinson {\it et al.} 1991]{RR}
Rowan-Robinson. M., {\it et al.} 1991, {\it Nature}, {\bf 351}, 719. 
	\bibitem[Rowan-Robinson 2001]{RR01}
Rowan-Robinson, M. 2001, \apj, {\bf 549}, 745.
	\bibitem[Sanders \& Mirabel 1996]{SM96} 
Sanders, D. B. \& Mirabel, I. F. 1996, \araa, {\bf 34}, 749.
	\bibitem[Sanders {\it et al.} 1988]{San88} 
Sanders, D., {\it et al.} 1988, \apj, {\bf 325}, 74.
	\bibitem[Sanders {\it et al.} 1989]{San89} 
Sanders, D., {\it et al.} 1989,	\apj, {\bf 347}, 29.
	\bibitem[Smith, Lonsdale \& Lonsdale 1998]{SLL98} 
Smith, H. E., Lonsdale, C., \& Lonsdale, C.
		1998, \apj, {\bf 492}, 137.
	\bibitem[Smith {\it et al.} 1998]{SLLD98} 
Smith, H. E., Lonsdale, C., Lonsdale, C. \& Diamond, P.
		1998, \apjl, {\bf 493}, L17.
	\bibitem[Smith {\it et al.} 2002]{Sm02} 
Smith, H. E., Lonsdale, C., Hurt, R. \& Siana, B.
		2002, \apj, {\it in preparation}.
	\bibitem[Stanford {\it et al.} 2001]{St01}
Stanford, S., {\it et al.} 2000, \apjs, {\bf 131}, 185.
	\bibitem[Steidel {\it et al.} 1999]{CS99}
Steidel, C., {\it et al.} 1999, \apj, {\bf 519}, 1.
	\bibitem[Tran {\it et al.} 2001]{Tr01}
Tran, Q., {\it et al.} 2001, \apj, {\bf 552}, 527.
	\bibitem[Wilkes {\it et al.} 2002]{W02}
Wilkes, B., {\it et al.} 2002, \apjl, {\it in press}. (astro-ph/0112433)
	\bibitem[Xu {\it et al.} 2001]{X01}
Xu, C., Lonsdale, C., Shupe, D., O'Linger, J., \&  Masci, F. 2001, 
	\apj, {\bf 562}, 179.
	\end{thebibliography}
\end{document}